\documentstyle[aps,twocolumn,prl,epsf]{revtex}

\newcommand{\pb}{\overline{P}}
\newcommand{\sg}{\sigma^2}
\newcommand{\lb}{\overline{\ell}}
\newcommand{\lbsq}{\overline{\ell^2}}
\newcommand{\bea}{\begin{eqnarray}}
\newcommand{\eea}{\end{eqnarray}}

\begin{document} 
\twocolumn[\hsize\textwidth\columnwidth\hsize\csname
@twocolumnfalse\endcsname

\title{Exact results and scaling properties of small-world networks} 
\author{R. V. Kulkarni\cite{e-mail}, E. Almaas and D. Stroud}
\address{Department of Physics, The Ohio State University,
         Columbus, OH 43210} 

\date{\today}

\maketitle

\begin{abstract} 
We study the distribution function for minimal paths in small-world
networks. Using properties of this distribution function, we derive
analytic results which greatly simplify the numerical calculation of
the average minimal distance, $\lb$, and its variance, $\sg$. We also
discuss the scaling properties of the distribution function. Finally,
we study the limit of large system sizes and obtain some analytic
results.
\end{abstract}

\draft 

\pacs{PACS numbers: 05.10.-a,05.40.-a,05.50.+q,87.18.Sn} 

\vskip1.5pc]

\par 

Recently Watts and Strogatz \cite{watts} have studied a class of
networks which can be `tuned' from an ordered to a random state by
varying a single parameter. For a range of parameter values, they
found that the networks resembled ordered networks locally but random
networks globally. They named this class of networks ``small-world''
networks because of their relevance to a well-known problem in
sociology\cite{mil}. Since their introduction, small-world networks
and their properties have received considerable attention,
\cite{pan,barth,barr1,mon,new1,barr2,men,kas,new2,kul,mou1,mou2,alb,barab,dm},
in part because of their possible application to a broad range of
systems, ranging from social networks\cite{mil} to coupled
oscillators\cite{kur}.

\par 

Much of the work characterizing the properties of small-world networks
has focused on the average minimal distance, $\lb$, separating two
randomly chosen points in the network. Previous work has shown that
$\lb$ has two scaling regimes: for small systems sizes $L$, it is
found that $\lb \sim L$, whereas for large $L$, $\lb \sim \log(L)$,
\cite{ball}. A scaling form for $\lb$ has been proposed and
numerically confirmed; however the nature of the crossover between the
two scaling regimes has been the subject of
debate\cite{barth,new1,men,new2,mou1,mou2}. In this work, we will
focus on some basic probability distributions for small-world
networks, and as a consequence obtain an {\em exact} result for $\lb$
which greatly aids its numerical computation.

\par 

We generate the networks following the prescription of Newman and
Watts \cite{new1}. We start with a 1-$d$ periodic lattice with $L=2 N$
sites and nearest neighbor connections. We then add short-cuts
uniformly with probability $p$ such that the average number of
short-cuts added is $x=p L$. We denote the distance between two sites,
counted along the lattice using only nearest neighbor links, as the
{\em Euclidean distance}. By contrast, the shortest distance between
two sites, counted along any bond including short-cut bonds, is called
the {\em minimal distance}.

\par

Using these definitions, we now introduce the following probability
functions: (i) $\pb(n|m)$, the probability that two sites are
separated by Euclidean distance $n$ given that their minimal distance
is $m$; (ii) $P(m|n)$, the probability that two sites have minimal
separation $m$ given that their Euclidean distance is $n$; and (iii)
$Q(m)$, the probability that two randomly chosen sites have a minimal
separation $m$. Recently, Dorogovtsev {\it et al} \cite{dm} have
introduced two exactly solvable models similar to small-world
networks. For these models they derive the explicit form of $P(m|n)$,
from which they obtain $\lb$ and other properties of their
networks. In this Letter, we derive the general form of $P(m|n)$ for
small-world networks, and we confirm it numerically. Using this form,
we derive an exact expression for $\lb$ and for the variance of $\lb$,
$\sg \equiv \lbsq-\lb^2$. We also study the scaling properties of
$P(m|n)$ and obtain some approximate results for it in the limit of
large $L$. Note that in describing the networks, we have considered
the case of coordination number of $2k=2$ for each site. However, our
arguments for the general form of $P(m|n)$ are valid for arbitrary
$k$. For simplicity we will consider the case $k=1$ in the rest of
this Letter and generalizations to arbitrary $k$ will be indicated as
appropriate.

\par 

We begin by deriving the general form of $\pb(n|m)$. First, since the
minimal distance cannot exceed the Euclidean distance, $\pb(n|m)=0$
for $n<m$. For $n>m$, the minimal path must use at least one
short-cut. But taking a short-cut is equivalent to randomizing the
position along the network, since the short-cuts are uniformly
distributed. Hence, $\pb(n|m)$ must be {\em independent} of $n$ for
all $n>m$. Finally, for $n=m$, it is not necessary to use any
short-cuts in the minimal path; so the arguments invoked for $n > m$
do not apply. Instead, $\pb(n|n)$ is determined by the constraint
that the probability distribution is normalized.

\par

We now derive the general form of $P(m|n)$. From elementary
probability theory, we have
\begin{equation}
\pb(n|m)*Q(m) = \left\{\begin{array}{ll}
			\frac{2}{L-1}P(m|n) & ~;~~n<N \label{pb}\\
			&	\\
                        \frac{1}{L-1}P(m|n) & ~;~~n=N
			\end{array}\right.
\end{equation}
From Eqn.\ (\ref{pb}) and the properties discussed in the previous
paragraph, $P(m<n|n) \equiv f(m)$ is {\em independent} of $n$, and
$P(m>n|n)=0$. Thus the general form of $P(m|n)$ is
\begin{equation}
P(m|n) = \Theta(n-m) f(m) + \bigl[1 - \sum_{m^\prime=1}^{n-1}f(m^\prime)\bigr] 
	\delta_{m,n}, \label{Pmn}
\end{equation}
\begin{figure}[tb]
\epsfysize=7.5cm
\centerline{\epsffile{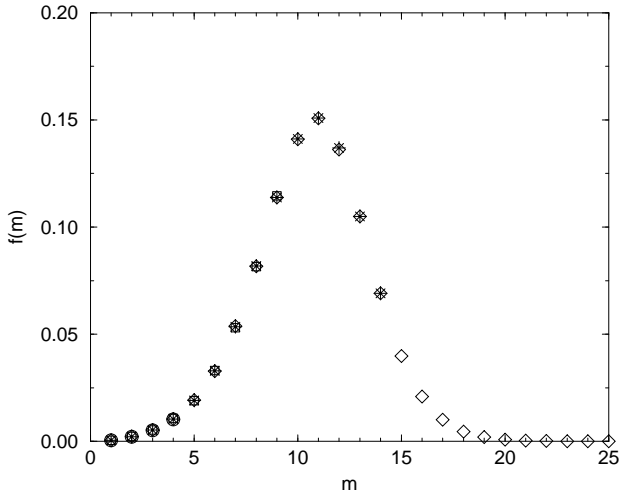}}
\caption{The values of $f(m)$ as obtained from $P(m|n)$, which is
calculated from simulations for the $n$-values $n=5$ ($\circ$), $n=10$
($\Box$), $n=15$ ($\ast$), and $n=500$ ($\diamond$). This figure
confirms the statement that $P(m\!<\!n|n)$ is independent of $n$. The
parameters for the figure are $L=1000$, $p=0.25$. Similar confirmation
has been obtained from simulations for a wide range of parameter
values.}
\label{fig1}
\end{figure}
\noindent
where $\Theta(x)$ is defined by $\Theta(x) = 0$ for $x \leq 0$ and
$\Theta(x) = 1$ for $x > 0$. We have numerically confirmed the
validity of this form, as shown in Fig.\ \ref{fig1}.

\par 

The fact that $P(m|n)$ is completely determined by $f(m)$ has some
surprising consequences, regardless of the exact form of $f(m)$. To
explore these consequences, we examine some other properties of
small-world networks. For example, besides $\lb$, the quantity
$<\!\!\ell(n)\!\!>$, which is the average minimal distance separating
two sites with Euclidean distance $n$, has been discussed in the
literature\cite{mou1,mou2}. We can express both these quantities in
terms of $P(m|n)$ as follows:
\begin{equation}
<\!\!\ell(n)\!\!>  =  \sum_{m=1}^{n} m P(m|n) \label{ln}
\end{equation}
\begin{equation}
\lb         =  \frac{1}{L-1} \Bigl[ 2 \sum_{n=1}^{N-1} 
	           \sum_{m=1}^{n}  m P(m|n)
		 + \sum_{m=1}^{N}  m P(m|N)\Bigr] \label{lbar1}  
\end{equation}
Similar expressions hold for $<\!\!\!\ell^2(n)\!\!\!>$,
$<\!\!\!\ell^3(n)\!\!\!>$, and $\lbsq$. Substituting the form of
$P(m|n)$ [Eqn.\ (\ref{Pmn})] into the expression for $\lb$ [Eqn.\
(\ref{lbar1})], we obtain
\begin{eqnarray}
\lb & = & \frac{1}{L-1}\biggl[2\sum_{n=1}^{N-1} \sum_{m=1}^{n-1} m f(m) + 
          2 \sum_{n=1}^{N}n\bigl[1 - \sum_{m=1}^{n-1}f(m)\bigr] \nonumber \\
    & + & \sum_{m=1}^{N-1} m f(m) + N \bigl[1 - \sum_{m=1}^{N-1}f(m)\bigr]
	\biggr]
\end{eqnarray}
which can be simplified to give the following exact expression:
\begin{figure}[tb]
\epsfysize=7.5cm
\centerline{\epsffile{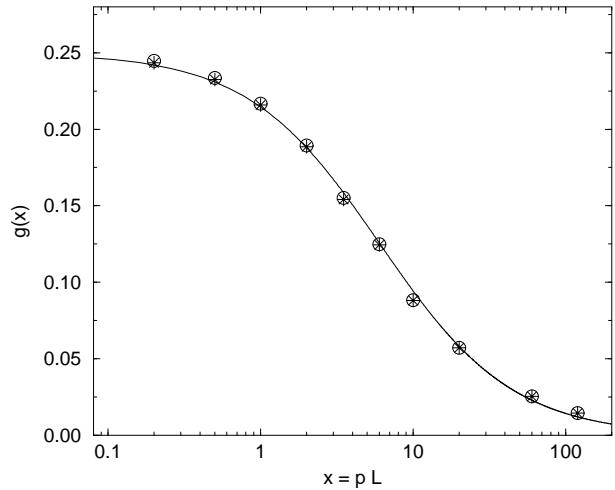}}
\caption{The average minimal separation, $\lb=L g(x)$, vs. average
number of short-cuts, $x= p L$, as obtained from (a) numerical
simulation by averaging over all pairs ($\ast$); (b) numerical
simulation using Eqn. \ref{lbar2} ($\circ$), and (c) Pade-fit as given
by Ref.$~[12]$ (solid line). This confirms the exact expression
Eqn. (\ref{lbar2}).}
\label{fig2}
\end{figure}
\begin{equation}
\lb ~=~ <\!\!\ell(N)\!\!>\left(1 + \frac{1}{L-1}\right) - 
	\frac{<\!\!\ell^2(N)\!\!>}{L-1}\label{lbar2}
\end{equation}
Similarly, we also obtain an exact expression for the variance of the
distribution of minimal distances, $\sg=\lbsq-\lb^2:$
\begin{eqnarray} 
\sg &=& <\!\ell^2(N)\!>(1+\frac{1}{L-1}) \nonumber \\
    &+& \frac{1}{3(L-1)}
	\Bigl[<\!\ell(N)\!>-4<\!\ell^3(N)\!>\Bigr] \nonumber \\
    &-& \Bigl[ <\!\ell(N)\!>(1+\frac{1}{L-1}) - 
	\frac{<\!\ell^2(N)\!>}{L -1} \Bigr]^2 \label{sigma}
\end{eqnarray} 
The surprising aspect of the above equations is that $\lb$ and $\sg$,
which are average properties of the entire network, are completely
determined by the mean separation of `diametrically opposite sites'
(d.o.s.), $<\!\!\ell(N)\!\!>$, and its higher moments
$<\!\!\ell^2(N)\!\!>$ and $<\!\!\ell^3(N)\!\!>$. This fact, in
particular, makes for significant numerical simplification in the
computation of $\lb$. Note that Eqns. (\ref{lbar2}) and (\ref{sigma})
can readily be generalized to any $k$ by performing the substitution
$L\mapsto \lceil L/k\rceil$.

\par

When the network has exactly one short-cut, we can calculate $\lb$
analytically using Eqn.\ (\ref{lbar2}). In this case, in the limit of
large $N$, we get $<\!\!\ell(N)\!\!>=\!\!\frac{2}{3}N$ and
$<\!\!\ell^2(N)\!\!>=\!\!\frac{1}{2}N^2$ which gives
$\lb=\frac{5}{12}N$. As expected, this is in perfect agreement with
the results obtained by Strang and Eriksson \cite{new2,strang}. We
have further confirmed Eqn. (\ref{lbar2}) by numerically computing
$\lb$ using the following two procedures: (i) Averaging the minimal
distance over {\em all} pairs of sites, and (ii) considering only
pairs of d.o.s. and
\begin{figure}[tb]
\epsfysize=7.5cm
\centerline{\epsffile{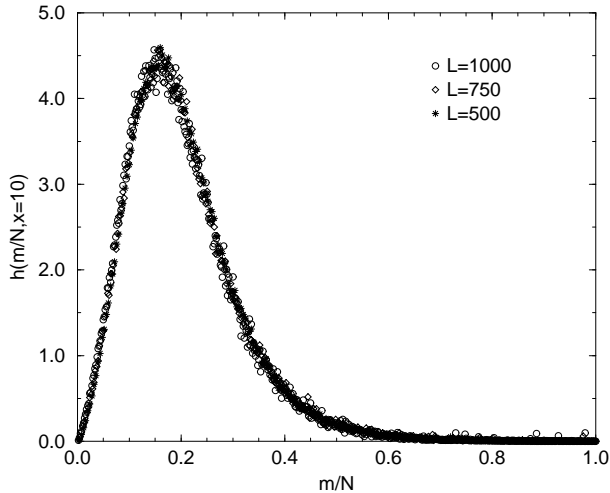}}
\caption{This figure confirms the proposed scaling form of $f(m,N,p)$
(Eqn. (\ref{fscal})) for $x=pL=10$ and system sizes: (a) $L=500$
($\ast$), (b) $L=750$ ($\diamond$), and (c) $L=1000$ ($\circ$). We
have confirmed this scaling collapse for a wide range of $x$-values.}
\label{fig3}
\end{figure}
\noindent
using Eqn. (\ref{lbar2}). The results, which are presented in
Fig. \ref{fig2}, indicate that the two procedures are equivalent.

\par 

The results obtained so far have been {\em independent} of the
functional form of $f(m)$. To gain further insight, we consider the
scaling properties of $f(m)$, following the real-space renormalization
group (RG) analysis of Newman and Watts \cite{new1}. This procedure
consists of blocking pairs of adjacent sites while preserving the
total number of short-cuts in the network. This gives for the
transformed lattice: $N'=N/2$ and $p'=2p$. We note the following
features of this transformation \cite{new1}: (i) the geometry of the
minimal paths is unchanged in almost all cases, and the number of
site-pairs for which the geometry does change is negligible for large
$L$ and small $p$, and (ii) the distance along the minimal path is
halved, i.e. $m'=m/2$ for large $L$ and small $p$. Now, by definition
$f(m)=P(m|N)~(m<N)$; thus to evaluate $f(m)$ we need to calculate the
average number of minimal paths of length $m$ between two d.o.s. in
the network. Furthermore we note that the RG-transformation maps two
pairs of d.o.s. into a single pair. This fact, in conjunction with
points (i) and (ii) above gives us
\begin{equation}
f'(\frac{m}{2},\frac{N}{2},2 p)=2 f(m,N,p) 
\end{equation}
For large $N$, taking the continuum limit, we can generalize the above
expression to
\begin{equation}
f'(\frac{m}{\lambda},\frac{N}{\lambda},\lambda p)=\lambda f(m,N,p) 
\end{equation}
These observations can now be summarized in the following scaling
form:
\newpage
\begin{eqnarray}
f(m,N,p) &=& \frac{1}{N} h(y,x)  \label{fscal}\\
\mbox{where} && y=\frac{m}{N},~~ x=2 p N  \nonumber
\end{eqnarray}
By fixing $x$, we have observed the scaling collapse of $f(m)$ for
different values of $N$ and $p$. This is demonstrated numerically for
$x=10$ in Fig.\ref{fig3}. Our simulations indicate that for any given
$x$, this scaling collapse holds for large enough $N$.

\par

It is interesting to note that the scaling properties of $\lb$ can be
derived from the scaling form of $f(m)$. Using the definition of $\lb$
(Eqn. (\ref{lbar1})) and the scaling form for $f(m)$, we get
\begin{eqnarray}
\lb   &= & \frac{L}{4} [1 - \int_0^1\!\!\!\!dy~(1-y)^2 h(y,x)] \\
      &= &  L~g(x) \label{lscal}
\end{eqnarray}
which is consistent with the scaling form proposed in previous works.
Similar scaling forms hold for $\lbsq$, $<\!\!\ell(N)\!\!>$, and
$<\!\!\ell^2(N)\!\!>$.

\par

\begin{figure}[tb]
\epsfysize=6.7cm
\centerline{\epsffile{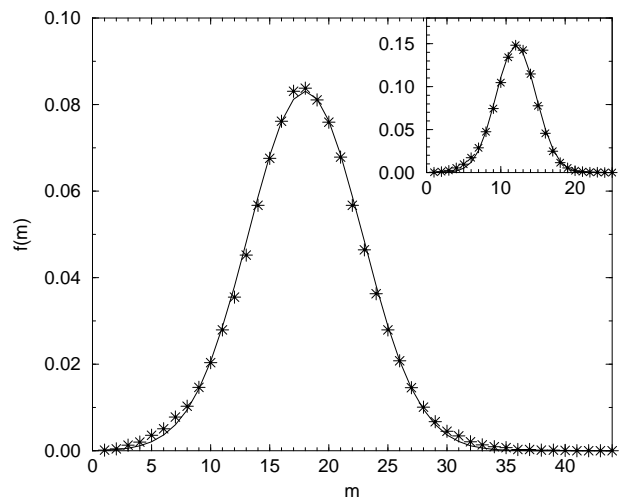}}
\caption{$f(m)$ vs. $m$ for $x=pL=250$ and $L=2000$. The solid line is
the gaussian fit to the calculated data. The inset shows the gaussian
fit for $x=500$ and $L=2000$. Note that with increasing $x$, the
gaussian becomes more sharply peaked.}
\label{fig4}
\end{figure}

We now consider the limit of large system sizes such that $x\gg1$. In
this limit, we have observed numerically that we can approximate
$f(m)$ by a gaussian distribution function:
\begin{equation}
f(m) = \frac{1}{\sqrt{2 \pi \sg_g}} e^{- \frac{(m- \mu_g)^2}{2 \sg}}
\label{gaus}
\end{equation}
where $\mu_g$ and $\sg_g$ are respectively the mean and variance of
the distribution. The corresponding fit for $x=250$ and $x=500$ is
shown in Fig. \ref{fig4}. Our simulations indicate that as $x$
increases, $(\mu_g/\sigma_g)$ also increases, as can be seen from the
figure.

\par

Using the gaussian approximation for $f(m)$, we are now able to
calculate the function $<\!\!\ell(n)\!\!>$, which has 
\begin{figure}[tb]
\epsfysize=7.5cm
\centerline{\epsffile{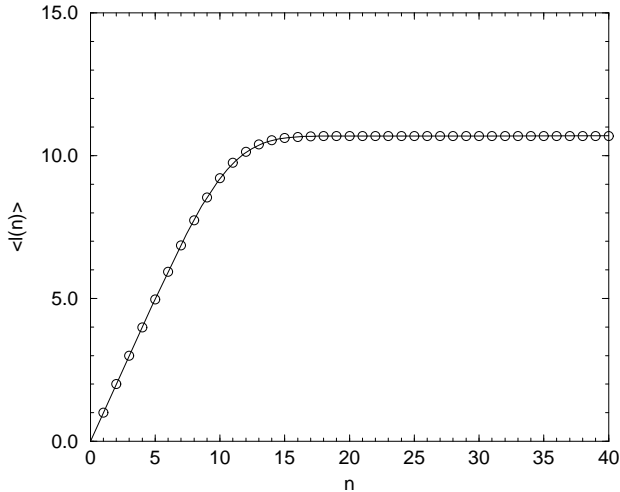}}
\caption{The mean distance $<\!\!\ell(n)\!\!>$ between two sites
having Euclidean separation $n$ for $x=pL=250$. Results are shown for:
(a) numerical simulation ($\circ$), and (b) analytic expression
(Eqn. (\ref{lofn})) (solid line). The analytic expression is an
excellent fit for $x\gg 1$.}
\label{fig5}
\end{figure}
\noindent
been discussed elsewhere \cite{mou1,mou2}. From Eqn. (\ref{ln}) and
(\ref{gaus}) in the limit $\mu_g \gg \sigma_g$, we obtain
\begin{eqnarray}
<\!\!\ell(n)\!\!> &=& n-\frac{1}{2}(n-\mu_g)\Bigl[\Phi\big(
	\frac{n- \mu_g}{\sqrt{2}\sigma_g}\bigr) + 1 \Bigr] \nonumber \\
      & &+ \frac{\sigma_g}{\sqrt{2 \pi}} e^{-\frac{(n- \mu_g)^2}{2 \sg_g}} 
	\label{lofn}
\end{eqnarray}
where we have assumed $\Phi(\mu_g/\sqrt{2}\sigma_g)=1$. In this limit,
substituting the above form of $f(m)$ into the definitions of
$<\!\!\ell(N)\!\!>$ and $<\!\!\ell^2(N)\!\!>$, we get
\begin{eqnarray}
\mu_g & = & <\!\!\ell(N)\!\!>   \\
\sg_g & = & <\!\!\ell^2(N)\!\!> - <\!\!\ell(N)\!\!>^2 
\end{eqnarray}
In particular, these equations imply that $\mu_g$ and $\sigma_g$ have
the following scaling forms: $\mu_g\sim L\, g_1(x)$ and $\sigma_g\sim
L\>g_2(x)$. Using these relations, we see that the gaussian ansatz for
$f(m)$ (Eqn. (\ref{gaus})) is consistent with the scaling form
proposed in Eqn. (\ref{fscal}). In Fig.$\,$\ref{fig5}, we compare
Eqn. (\ref{lofn}) to results from our simulations for $x=250$. In the
limit $L\to\infty$, we have $(\sigma_g/\mu_g) \to 0$, which upon
substitution into Eqn. (\ref{lofn}) gives us
\begin{equation}
<\!\!\ell(n)\!\!> = \left\{ \begin{array}{ll}
		n                  &~;~~n <~   <\!\!\ell(N)\!\!> \\
		<\!\!\ell(N)\!\!>  &~;~~n \ge~ <\!\!\ell(N)\!\!>
		            \end{array}
		\right.
\end{equation}
This expression for $<\!\!\ell(n)\!\!>$ is consistent with that
derived in Ref. \cite{mou1,mou2} in the same limit.

\par 

In conclusion, we have studied the probability distribution for
minimal path-lengths in small-world networks. We have presented
arguments for the general analytical form this distribution must
take. Using this form, we have derived some exact relations which will
significantly aid computational efforts. We have obtained an
approximate scaling form for this probability distribution in the
limit of large system sizes. It is our hope that further efforts along
these lines will provide a better understanding of the structure of
small-world networks.

\par

This work has been supported by NASA through Grant NCC8-152 and NSF
through Grant DMR97-31511. Computational support was provided by the
Ohio Supercomputer Center, the San Diego Supercomputer Center, and the
Norwegian University of Science and Technology (NTNU).

\end{document}